\documentclass[aps,pra,twocolumn,groupedaddress,showpacs]{revtex4}
\usepackage{graphicx}

\usepackage{ulem}

\begin{document}

\title{Antiphase dynamics in a multimode semiconductor laser with optical injection}

\author{S. Osborne$^{1}$, A. Amann$^{1}$, K. Buckley$^{1}$, G. Ryan$^{1}$, S. P. Hegarty$^{1}$, 
G. Huyet$^{1,2}$ and S. O'Brien$^{1}$}

\affiliation{$^{1}$Tyndall National Institute, University College, Lee Maltings, Cork, Ireland \\
$^{2}$Department of Applied Physics and Instrumentation, Cork Institute of Technology, Cork, Ireland}

\begin{abstract}
A detailed experimental study of antiphase dynamics in a two-mode 
semiconductor laser with optical injection is presented. The device 
is a specially designed Fabry-P\'erot laser that supports two primary 
modes with a THz frequency spacing. Injection in one of the primary 
modes of the device leads to a rich variety of single and two-mode 
dynamical scenarios, which are reproduced with remarkable accuracy 
by a four dimensional rate equation model. Numerical bifurcation 
analysis reveals the importance of torus bifurcations in mediating 
transitions to antiphase dynamics and of saddle-node of limit cycle 
bifurcations in switching of the dynamics between single and two-mode 
regimes.  
\end{abstract}


\maketitle

\section{Introduction}

The semiconductor laser with optical injection is a conceptually simple 
system that nevertheless has complex nonlinear dynamical properties. 
Phenomena of interest that have been observed experimentally include 
subharmonic resonance \cite{varangis_97}, period doubling route to chaos 
\cite{kovanis_95, gatare_06} and bistability of wave-mixing and injection 
locked states \cite{hui_91, huang_99}. Use of numerical continuation tools has enabled 
diverse dynamical states of the single mode edge-emitting system with 
optical injection to be linked together in a consistent way \cite{wieczorek_99}. 
More recently, a global picture of the complex dynamics of polarization 
modes in VCSELs with orthogonal optical injection has also begun to emerge 
\cite{gatare_07}.  

For both fundamental and applied reasons, interest in multimode 
dynamics of semiconductor lasers has increased. This follows from their 
increased dynamical complexity, as well as possible applications in optical 
signal processing \cite{tsimring_96, takenaka_05}. Multiple oscillating modes allow for 
antiphase behaviour, where mutual coupling of individual modal degrees of freedom 
leads to anticorrelated intensity dynamics \cite{wiesenfeld_90, 
khandokhin_98, larotonda_99, uchida_01, yacomotti_04}. 

Recently, we have developed a specially engineered two-mode diode laser 
with a large (THz) primary mode spacing \cite{obrien_06}. The device 
can be biased such that it oscillates on two modes simultaneously with the 
same average power level. Unlike the case of polarization
modes in VCSELs, the mode spacing is in the highly non-degenerate regime,
allowing us to develop studies of the multimode nonlinear dynamics
in the minimal possible two mode system. These studies provide a means 
to test models of the multimode diode laser and to reveal new bifurcation 
structure that occurs beyond the single mode approximation \cite{osborne_08}.   

A number of approaches have been developed to modeling the multimode 
dynamics of edge-emitting semiconductor lasers. Partial differential 
equations will generally reproduce experimental results, but such an 
approach suffers from computational difficulty and the reduced level
of physical insight provided \cite{white_98, huyet_99}. Rate equation 
approaches have proven very successful in the case of modeling the 
dynamical response of single mode semiconductor lasers. In the case of 
multimode dynamics, however, a number of different models have been 
developed that differ in how the dynamics of the carrier density 
are described \cite{buldu_02, masoller_05, koryukin_04, yacomotti_04}.  

Here we provide an experimental and theoretical study of antiphase dynamics when one 
of the two primary modes of the device is optically injected. Among the rich variety 
of multimode dynamical phenomena we have found limit-cycle, quasiperiodic 
and chaotic antiphase dynamics as well as regions where the optical power is largely 
switched to the \textit{uninjected} mode. As we will show, all of these dynamics, as 
well as the familiar injection locking on a single mode, can be found by performing a sweep 
over a large detuning range of the injected field for a certain fixed injected field 
intensity. As the central theme of this study, we provide a detailed description of 
regions of quasiperiodic antiphase dynamics using spectrally resolved power spectral 
measurements and intensity time traces. We then demonstrate that a four dimensional 
rate equation model reproduces the overall experimental picture very well. In particular, 
we show how regions of quasiperiodic antiphase dynamics that we have observed experimentally 
are reproduced in numerical simulations and that new bifurcation structure, which is 
particular to the two mode system considered, governs the appearance of these dynamics. 

This paper is organised as follows. In section \ref{sec:experiment}, we describe our 
device and experimental set-up and we provide optical and power spectral data that illustrate 
the variety of dynamical scenarios that can be observed as a function of frequency 
detuning at a fixed injected field strength. We also provide intensity time traces 
and power spectral data characterizing two regions of quasiperiodic antiphase dynamics. 
In section \ref{sec:model-device-resp}, we introduce our model equations that describe 
the two-mode injected system, showing how the overall experimental picture as well as 
specific examples of antiphase dynamics are qualitatively reproduced. Finally, in section 
\ref{sec:bifurcation-analysis}, we describe the bifurcation structure that leads to the 
antiphase dynamics observed. 

\section{EXPERIMENT}
\label{sec:experiment}

\begin{figure}
\includegraphics[width=0.9\columnwidth]{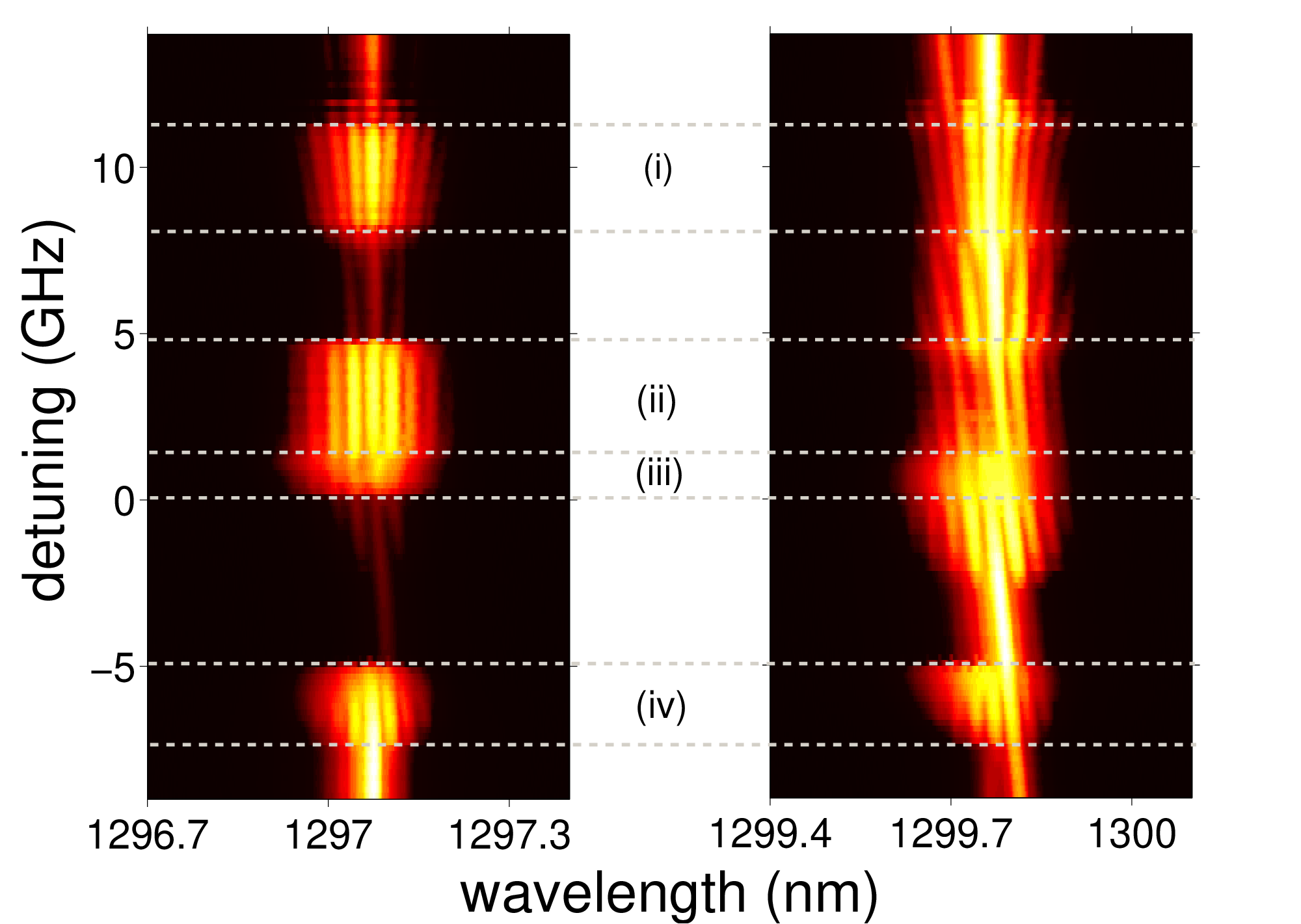}
\caption{\label{pra1} (Color online) Optical spectra of the two-colour device as the frequency detuning, 
$\Delta \omega$ is varied at a fixed injection. Left: uninjected mode, $\nu_{1}$. Right:
Injected mode, $\nu_{2}$.}
\end{figure}

For the primary mode spacing of four fundamental FP modes (480 GHz), we
observe simultaneous lasing of the two primary modes. We adjust the device current 
in order that the time averaged optical power in each primary mode of the free running
laser is equal. Details of the design and free running lasing characteristics of the 
device we consider can be found in Ref \cite{obrien_06}.

For our experiment, we use the two-colour laser in a master-slave configuration, where the 
master laser is a tunable laser with $< 100$ kHz linewidth. Choosing a fixed injected power 
level, we vary the detuning, $\Delta \omega$, of the injected field from the long wavelength 
mode of the device, $\nu_{2}$. The optical spectra of the two primary modes are shown as a 
function of the frequency detuning of the injected signal in Fig. \ref{pra1} for a detuning range of 
$-9 \mbox{ GHz} < \Delta \omega < 14 \mbox{ GHz}$. Despite the fact that we vary only the 
frequency detuning, Fig. \ref{pra1} indicates that we are dealing with a rich dynamical 
scenario; regions of multiwavelength dynamics are interspersed with regions of single or 
nearly-single mode dynamics, where the optical power can be concentrated in either of the 
primary modes. In sections \ref{sec:model-device-resp} and \ref{sec:bifurcation-analysis}
we will show that a minimal extension of the single mode rate equation model by a single
dynamical variable  is sufficient to account for the presence of the uninjected mode and 
reproduces the experimental data with remarkable accuracy.  

\begin{figure}
\includegraphics[width=0.9\columnwidth]{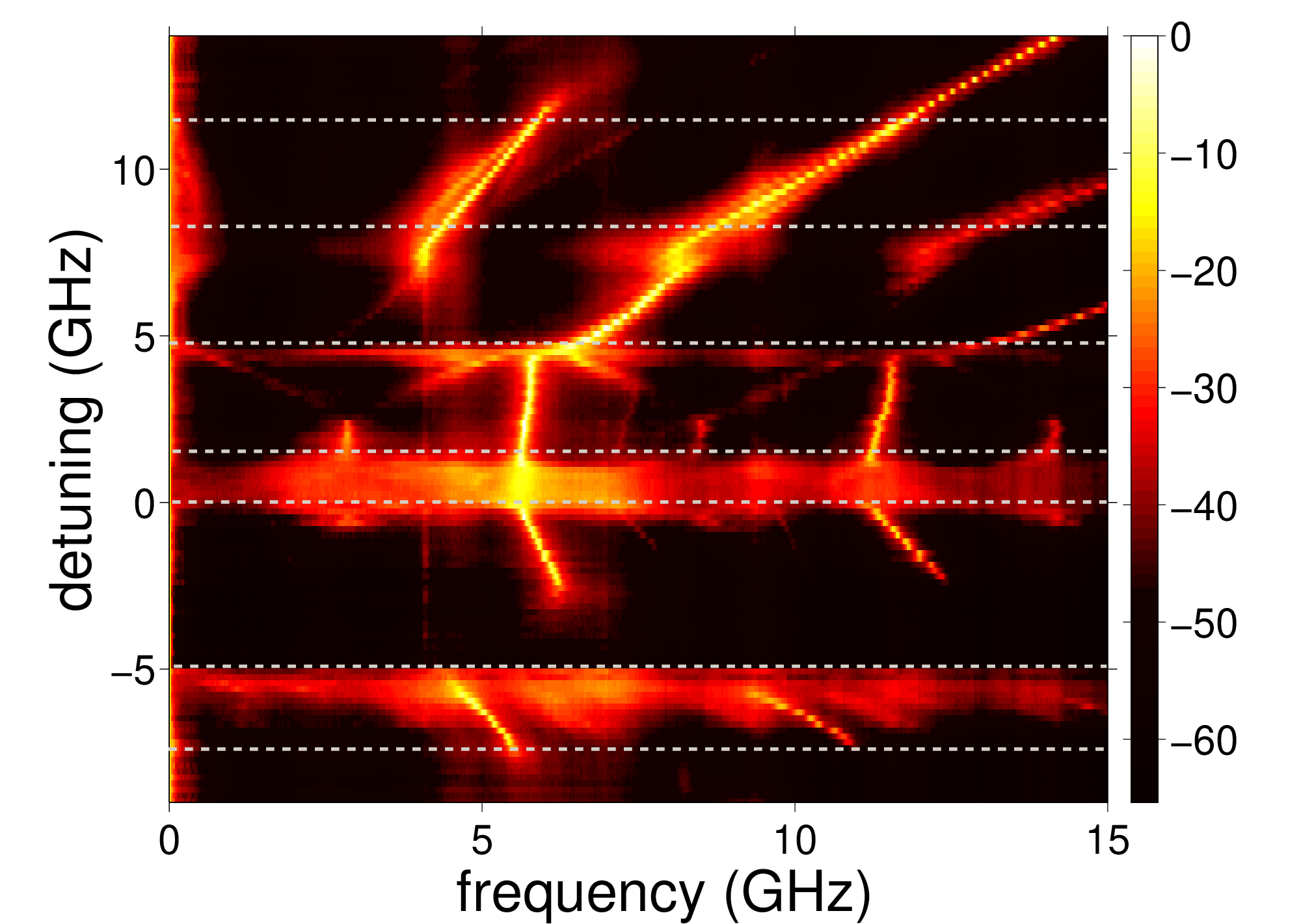}
\caption{\label{pra2} (Color online) Power spectral density of the total intensity as a function of 
frequency detuning. The injected power level is the same as in Fig. \ref{pra1}}
\end{figure}

For this injected power level the power spectral density of the total intensity is plotted in Fig.
\ref{pra2}. By considering the spectrum of the total intensity we naturally exclude the antiphase
frequency components that are present in each region of two mode dynamics. From Figs \ref{pra1} 
and \ref{pra2}, the regions of stable and unstable injection locking where the 
uninjected mode is off can be identified near zero detuning. In the injection locking region, 
a sharp single peak at the frequency of the injected field is present in the optical spectrum at 
long wavelength. In the region of unstable locking, which extends from approximately zero 
detuning to -2.5 GHz, a single frequency that is associated with undamped relaxation oscillations 
is visible in Fig. \ref{pra2}. This frequency is almost constant at c. 5.5 GHz, which we  
identify as the relaxation oscillation frequency of the device.

At large positive frequency detuning of 14 GHz, we find a broad region of nearly 
single mode dynamics, where the optical power is concentrated in the injected mode. 
In this region the injected field and two weaker symmetrically placed components
can be clearly resolved in the optical spectrum at long wavelength. Examining Fig. 
\ref{pra2}, we find that in the corresponding region only the detuning frequency is 
present. This is then a region of wave mixing or beating between the injected field 
and the long wavelength primary mode. 

\begin{figure}
\includegraphics[width=0.9\columnwidth]{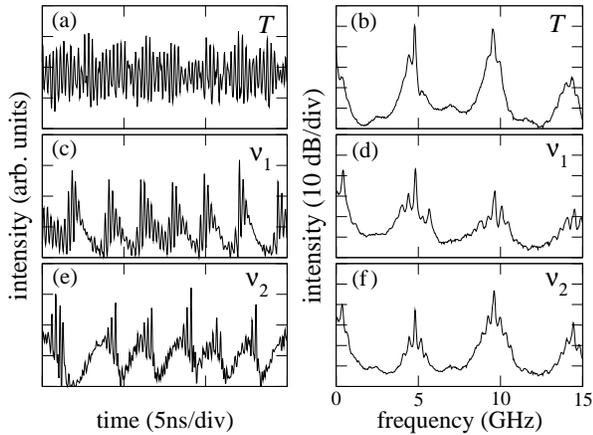}
\caption{\label{pra3} Experimentally measured antiphase dynamics, region (i). The frequency detuning
is 10 GHz and the injected power level is the same as in Fig. \ref{pra1}. Left panels: 
Intensity time traces. Right panels: Power spectral densities. (a) and (b): Total Intensity. 
(c) and (d): uninjected mode. (e) and (f): Injected mode.}
\end{figure}

As the detuning is decreased, a period doubling bifurcation occurs and we enter a region where 
a new frequency equal to one-half of the detuning frequency  is present in Fig. \ref{pra2}. 
As indicated in Fig. \ref{pra1}, within this broad region we find region (i) of two mode 
dynamics, where the optical spectra of both modes are are considerably broadened. 
A representative set of power spectral densities and time traces of the total intensity 
and the individual modal intensities in region (i) are shown in Fig. \ref{pra3}. In this 
figure the detuning is 10 GHz. As expected the power spectrum of the total intensity shown 
in Fig \ref{pra3} (b) comprises two broadened peaks which include distinctive shoulders. 

In the power spectra of the intensities of the individual modes shown in Fig. \ref{pra3} (d) and (f) we 
can see a strong antiphase frequency component with a distinct peak at c. 400 MHz. We see also that the 
features associated with the injection frequency and its subharmonic in each of these figures are broadened 
further because of the formation of satellites due to mixing with the low frequency antiphase component. 
Note also that the detuning of the injected signal is the dominant frequency in the dynamics of the 
injected mode [Fig. \ref{pra3} (f)] but the subharmonic near the relaxation oscillation frequency 
dominates in the dynamics of the uninjected mode [Fig. \ref{pra3} (d)]. 

We have confirmed that the low antiphase frequency is an independent frequency component that 
is not a linear combination of the other frequencies present. The low frequency is clearly visible 
as an envelope modulation of the intensity in the time traces of the individual modes shown in 
Fig. \ref{pra3} (c) and (e). These antiphase time traces have a distinctive sawtooth structure that 
results from the growing or decaying oscillations that form each pulse. They are  
remarkably similar to the so-called regular pulse packages that can be observed in 
semiconductor lasers with optical feedback in the short external cavity regime \cite{heil_01}. 
However, in the case of optical injection, the fast time scale is determined by the relaxation
oscillation and its harmonics, rather than the external cavity frequency. 

As the detuning is decreased further a reverse period-doubling occurs and a simple single mode wave 
mixing region is found. Adjacent to this region, we next find a second broad region of two mode dynamics, 
which has a lower boundary at zero detuning. For reasons that will become clear, we subdivide this 
region into two, and label the sub-regions as regions (ii) and (iii). Broadly speaking, region (ii) 
is found once the detuning is less than the relaxation oscillation frequency. Note that across the 
boundary between the single mode wave mixing region and region (ii), the power spectrum of the total 
intensity briefly becomes strongly broadened. Then, as the detuning is decreased further, the dynamics 
become simpler with the relaxation oscillation frequency, and its harmonics, largely determining the 
dynamics of the total intensity. 

\begin{figure}
\includegraphics[width=0.9\columnwidth]{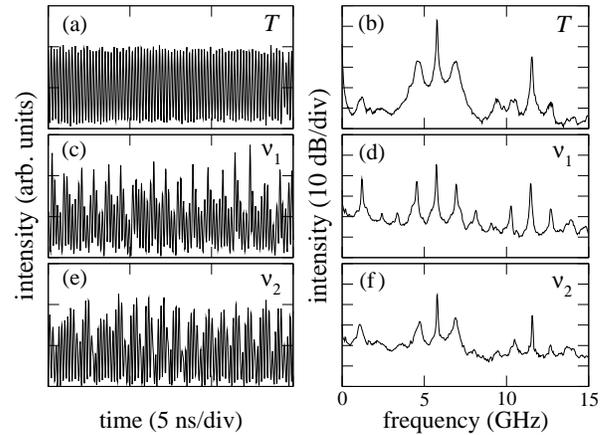}
\caption{\label{pra4} Experimentally measured antiphase dynamics, region (ii). The frequency detuning
is 4 GHz and the injected power level is the same as in Fig. \ref{pra1}. Left panels: 
Intensity time traces. Right panels: Power spectral densities. (a) and (b): Total Intensity. 
(c) and (d): uninjected mode. (e) and (f): Injected mode.}
\end{figure}

A representative set of power spectral densities and time traces in region (ii) from inside this 
boundary are shown in Fig. \ref{pra4}. The dynamics of the total intensity is now
largely determined by the relaxation oscillation and its harmonic although  weak and 
broadened satellite features are also visible [cf. Fig. \ref{pra4} (b)]. 

From Fig. \ref{pra4} (d) and (f), we see that the dynamics of the individual modes 
are determined by a series of frequencies. In particular, there is a strong low frequency 
component present in the individual dynamics of both modes that is almost completely
in antiphase. As was the case in region (i), this low frequency component is  
responsible for the appearance of a cascade of frequencies that are centered at the 
relaxation oscillation peak and its harmonic. 

As the detuning is decreased further, the frequency difference between the relaxation oscillation 
and the detuning approaches half of the relaxation oscillation frequency. In this region a weak 
period two region of the total intensity is found. For detunings less than approximately one-half 
the relaxation oscillation frequency we find coupled chaotic dynamics that are characterised by 
broad-band spectra. 

For negative detunings, the single mode locking region is found, which extends
to a detuning of approximately -5 GHz. At the boundary of the stable locking region, inside of 
which  there is no structure in the power spectrum, we find region (iv), where the spectrum is 
again strongly broadened. Decreasing the detuning further, the spectrum is almost structureless. 
Here, a weak feature at the detuning frequency can be seen. This corresponds to the transfer 
of the optical power to the uninjected mode. 

\section{MODELING OF THE DEVICE RESPONSE}
\label{sec:model-device-resp}

We have adapted the well known model of a single mode laser \cite{buldu_02, heil_01} with 
optical injection to account for the presence of a second lasing mode. The system of equations 
in normalized units may be written as follows \cite{osborne_08}:
\begin{eqnarray} 
&\dot{E}_{1} =  \frac{1}{2} (1 + i\alpha)(g_{1}(2n + 1) - 1)E_{1} \label{eq:1}\\
&\dot{E}_{2} =  \left[\frac{1}{2} (1 + i\alpha)(g_{2}(2n + 1) - 1) - i\Delta\omega\right]E_{2} + K  \label{eq:2}\\
&T\dot{n} = P - n - (1 + 2n)\sum_{m}g_{m}|E_{m}|^{2} \label{eq:3} 
\end{eqnarray}
where the nonlinear modal gain is 
\begin{equation}
g_{m} = g_{m}^{(0)}\left(1 + \epsilon\sum_{n}\beta_{mn}|E_{n}|^{2}\right)^{-1}.
\end{equation}

Here $E_{1}$ and $E_{2}$ are the normalized complex electric field amplitudes of the modes and  $n$
is the normalized excess carrier density. The bifurcation parameters are the normalized injected field 
strength $K$ and the angular frequency detuning $\Delta\omega$. Further parameters are the phase-amplitude 
coupling $\alpha$, the product of the carrier lifetime and the cavity decay rate $T$, the normalized 
pump current $P$, and the linear modal gain $g_{m}^{(0)}$. In our numerical simulation we used the values
$\alpha = 2.6$, $T^{-1} = 0.00125$, $P = 0.5$, (twice threshold), and $g_{m}^{(0)} = 1$. Then the value of the 
relaxation oscillation frequency is $\omega_{\text{RO}}=\sqrt{2P/T} \sim 5.5$ GHz, in agreement with 
experiment. The cross and self saturation are determined by $\epsilon\beta_{mn}$ and we use the values 
$\epsilon = 0.01$, $\beta_{12} = \beta_{21} = 2/3$ and $\beta_{11} = \beta_{22} = 1$, which is 
consistent with the stability of the two mode solution in the free running laser.

Note that although we have provided a complex equation for the field $E_{1}$, the phase of $E_{1}$ 
is in fact decoupled leading to a four-dimensional system of equations where only the intensity of 
the uninjected mode influences the dynamics. In addition, the single mode dynamics is contained 
within the invariant submanifold ($E_{1} = 0$) in these equations. Our model equations are therefore 
a minimal extension of the (three dimensional) single mode case and follow from the fact that the 
primary mode spacing in the device considered is in the highly non-degenerate regime.  

This system only considers a single averaged carrier density and a general cross and self saturation 
of the gain. More complex models of the multiwavelength dynamics are available that are derived 
by considering additional Fourier components of the carrier density. One example includes the 
effect of static spatial hole burning explicitly leading to a system with two carrier density 
equations, one defined for each field \cite{koryukin_04}. However, we have found that 
these equations lead to some unphysical results in the injected system, such as a stable injection 
locked region where the uninjected mode is not suppressed. 

Another modeling approach \cite{mandel_93, yacomotti_04} includes an asymmetric term that is 
associated with a dynamic grating formed in the carrier density profile. This term effectively 
represents the mutual injection or wave mixing interaction of the primary modes and has been 
shown to play a role in the \textit{free running} switching dynamics of Fabry-P\'erot lasers 
\cite{yacomotti_04}. This grating is also responsible for the formation of four-wave mixing 
sidebands in the free running two-mode laser considered here \cite{obrien_06}. However, we have 
found equivalent experimental results for injection at both the long and short wavelength primary 
modes, suggesting that this grating does not significantly affect the dynamics.
We attribute this to the very large separation of the primary modes of the device (480 GHz), which 
means that the dynamic grating is weakly developed, with large sidebands observed because of 
enhancement by the Fabry-P\'erot cavity.

\begin{figure}
\includegraphics[width=0.9\columnwidth]{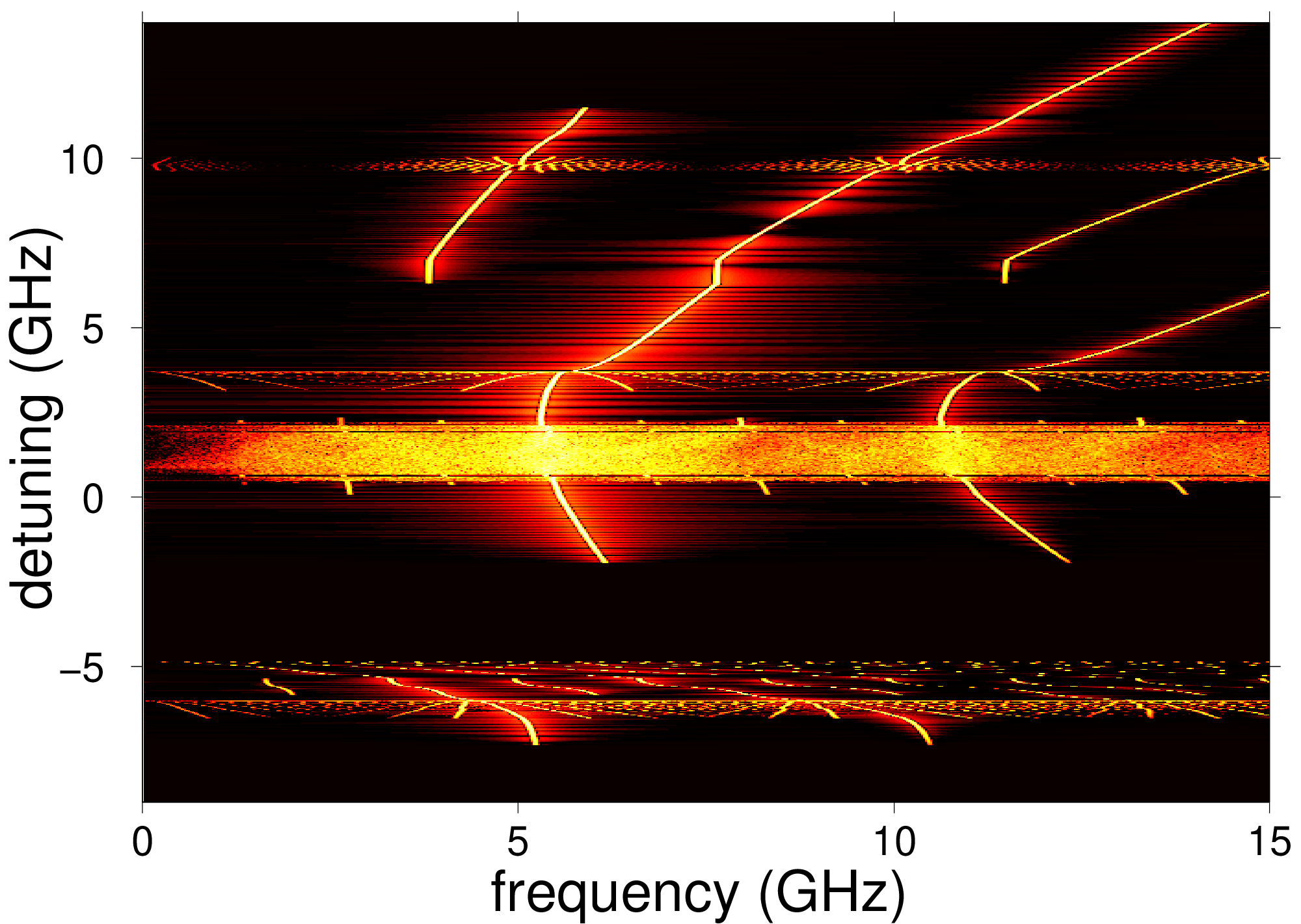}
\caption{\label{pra5a} (Color online) Power spectral density  of the total field intensity 
$|E_1|^2 + |E_2|^2$ obtained from numerical integration of Eqs. (\ref{eq:1})--
(\ref{eq:3}) as a function of the detuning. The injected field strength is 
$K = 0.008$.}
\end{figure}

Using the model equations (\ref{eq:1})--(\ref{eq:3}), in Fig. \ref{pra5a} we plot the 
power spectral density  of the total field intensity $|E_1|^2 + |E_2|^2$ 
for fixed $K$ and for detuning $-9 \mbox{ GHz} < \Delta \omega < 14 \mbox{ GHz}$.
The value of $K$ is 0.008 which was chosen as it provided the best possible agreement 
with experiment over the whole range of  frequency detuning. When compared with the 
experimental result of Fig. \ref{pra2}, we find that the level of qualitative agreement 
is excellent.

\begin{figure}
\includegraphics[width=0.9\columnwidth]{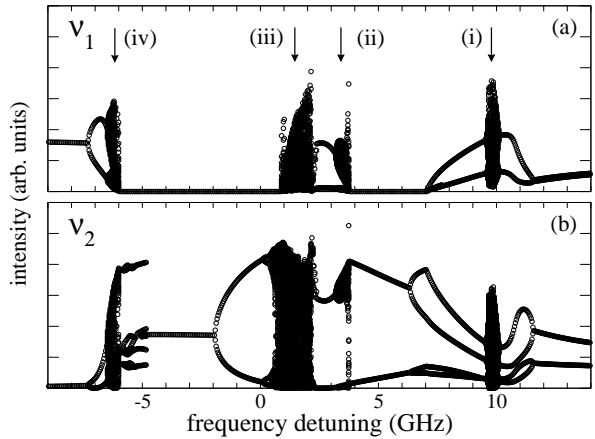}
\caption{\label{pra5b}  Local extrema of the field intensities (a) $|E_1|^2$ and
(b) $|E_2|^2$ obtained from numerical integration of Eqs. (\ref{eq:1})--(\ref{eq:3}) 
as a function of the detuning. The injected field strength is $K = 0.008$. Upper
panel: uninjected field. Lower panel: Injected field. }
\end{figure}

To further illustrate that the individual modal dynamics are also well reproduced, 
in Fig. \ref{pra5b} we plot the local extrema of the numerically calculated field 
intensities, $|E_1|^2$ and $|E_2|^2$, over the same detuning range. 

If we examine the structure of Fig. \ref{pra5b}, we can immediately identify two 
regions where the uninjected mode is off and that the position and extent of 
these regions is in good qualitative agreement with the experimental results shown 
in Fig. \ref{pra1}. Thus not only do the model equations reproduce the dynamics
of the total intensity, they also provide an accurate picture of the regimes of
single and multimode dynamics of the system. Note that this agreement extends
to the model reproducing the strong suppression of the injected mode at large 
negative detunings that is observed experimentally. 

In Fig.  \ref{pra5b}, vertical arrows indicate the regions (i) to (iv) of 
two mode dynamics that we have identified experimentally. For detailed comparison 
with experiment, we now provide the numerical intensity time traces 
and corresponding power spectra for the two regions of antiphase dynamics 
labelled (i) and (ii). 

\begin{figure}
\includegraphics[width=0.9\columnwidth]{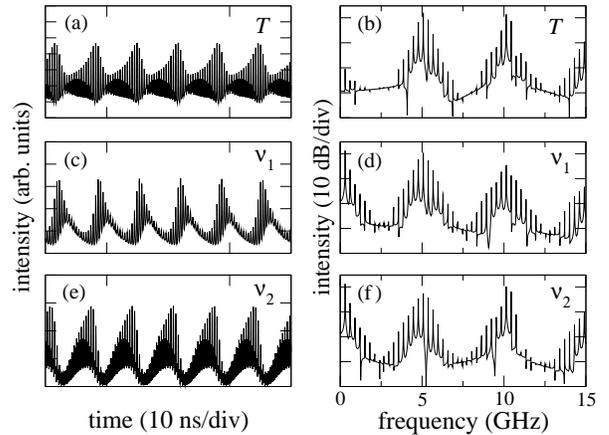}
\caption{\label{pra6} Numerically computed antiphase dynamics, region (i). The frequency 
detuning is 10.0 GHz. The injected field strength is $K = 0.008$. Left panels: 
Intensity time traces. Right panels: Power spectral densities. (a) and (b): Total Intensity. 
(c) and (d): uninjected mode. (e) and (f): Injected mode.}
\vspace{0.75cm}
\end{figure}

In Fig. \ref{pra6}, we show these data for the total intensity and 
for the intensities of each of the two modes at a detuning of 10.0 GHz, which
is located in region (i). What is most striking is the reproduction of 
the sawtooth pulse package structure that was seen experimentally. The envelope 
period of the pulse packages is seen to be c. 300 MHz, in broad agreement with 
experiment. The cascade of equally spaced frequencies that begins at this 
envelope frequency are also almost completely in antiphase. The power spectrum 
of the total intensity is also seen to comprise two broad cascades formed by 
the envelope frequency, one centered at the detuning frequency and another at 
the relaxation oscillation frequency. However, the number of these peaks is much 
larger than was found experimentally in the case of the individual modes and, 
in fact, the cascade structure that is seen in the numerical data for the total 
intensity is poorly resolved in the experiment.  
 
\begin{figure}
\includegraphics[width=0.9\columnwidth]{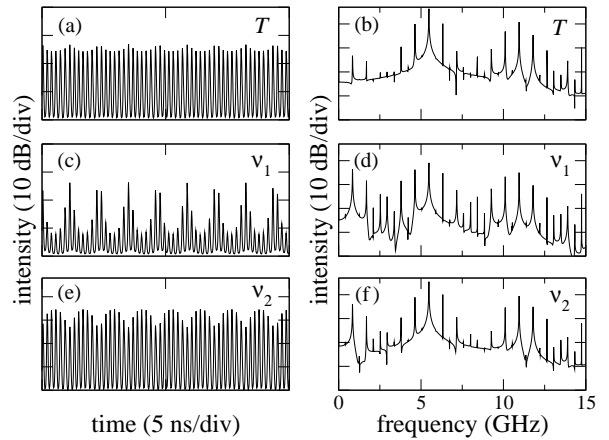}
\caption{\label{pra7} Numerically computed antiphase dynamics, region (ii). The frequency 
detuning is 3.4 GHz. The injected field strength is $K = 0.008$. Left panels: 
Intensity time traces. Right panels: Power spectral densities. (a) and (b): Total 
Intensity. (c) and (d): uninjected mode. (e) and (f): Injected mode.}
\end{figure}

In Fig. \ref{pra7}, we show these data for the total intensity and for the 
intensities of each of the two modes at a detuning of 3.4 GHz, which
is located in region (ii). We have had to chose a slightly smaller value for 
the detuning frequency for this simulation because of the discrepancy that 
exists between the boundary of the central two mode region that was found 
numerically and in the experiment. Nevertheless, the qualitative agreement 
with the experimental time traces and power spectra is again very good in 
this region. We note in particular the experimental intensity waveforms of 
the individual modes that are peaked in the case of the uninjected mode 
and rounded for the injected mode are well reproduced. 

\section{BIFURCATION ANALYSIS}
\label{sec:bifurcation-analysis}

\begin{figure}
\includegraphics[width=0.9\columnwidth]{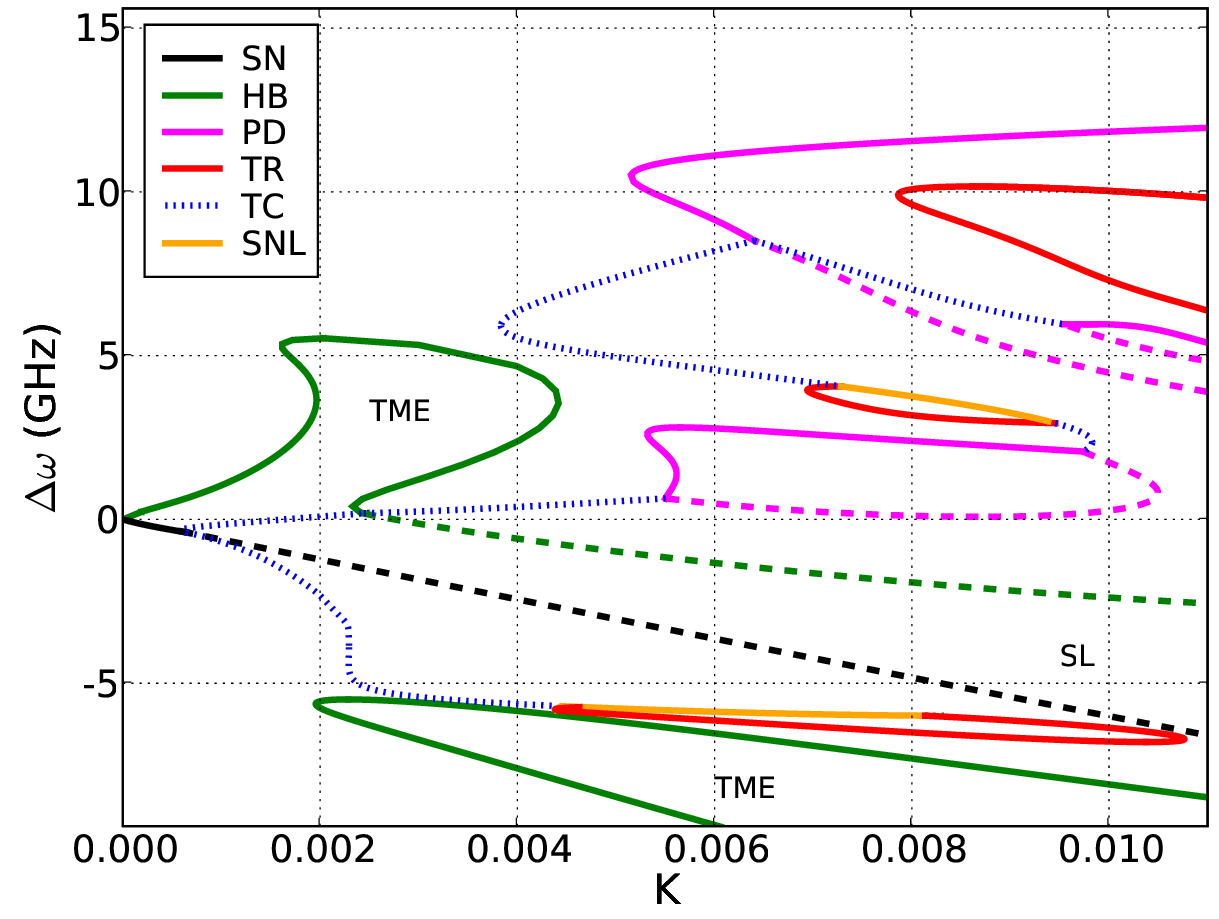}
\caption{\label{pra8} (Color online) Bifurcation diagram in the $\Delta \omega$
  vs. $K$ plane of Eq.~(\ref{eq:1})--(\ref{eq:3}) calculated using the
  numerical continuation tool AUTO-07p \cite{auto}.  Solid and dashed
  lines denote bifurcations of two mode and single mode states,
  respectively. The different types of bifurcations are indicated by
  \textit{SN} (saddle node), \textit{HB} (Hopf), \textit{PD} (period
  doubling), \textit{SNL} (saddle node of limit cycle), and
  \textit{TR} (torus). The striped blue lines \textit{TC} denote
  transcritical bifurcations between stable single mode and stable two
  mode states.  The single mode (stable) locking and two mode equilibrium 
  regions are denoted by SL and TME, respectively. Only bifurcation lines
  which affect stable states are shown.}
\end{figure}

The bifurcations evident from Fig.~\ref{pra5a}  and Fig.~\ref{pra5b} can be
understood from the global bifurcation diagram in the $\Delta \omega$
vs.\ $K$ plane as shown in Fig.~\ref{pra8}. Let us focus on the
vertical line at injection strength $K=0.008$, which was used in
Fig.~\ref{pra5a}  and Fig.~\ref{pra5b} and identified to best fit the
experimental data shown in Figs.~\ref{pra1} and \ref{pra2}.  At large
$\Delta \omega =15\text{ GHz}$ we are in a two mode wave mixing region,
with the uninjected mode $\nu_1$ largely suppressed and weakly modulated.
This agrees with the experimental observation of
a weak but non-vanishing uninjected mode power in the region above region
(i) in Fig.~\ref{pra1}. As we lower the detuning frequency to $\Delta
\omega\approx 12\text{ GHz}$, which is about twice the relaxation
frequency $\omega_{\text{RO}}$, the dynamical coupling between the
uninjected mode and injected mode via the carrier density $n$
[Eq. (\ref{eq:3})] becomes relevant, and gives rise to a two mode 
period doubling bifurcation (magenta line in Fig.~\ref{pra8}). This
bifurcation line forms the upper boundary of the dynamical two mode
region (i), which we experimentally identified in Figs.~\ref{pra1} and
\ref{pra2}. 

As we further decrease $\Delta \omega$, we cut through a
regime which is delineated by a torus bifurcation line (red line in
Fig.~\ref{pra8}) at $\Delta \omega\approx 10\text{ GHz}$. We emphasize
that this torus bifurcation is a feature the two mode system that
is not present in the single mode case. Due to this torus bifurcation an
additional incommensurate low frequency of about  $300 \text{ MHz}$
appears as previously identified in the discussion of the
characteristic sawtooth antiphase time traces of Fig.~\ref{pra6}.
Note that outside of the boundary of the  torus bifurcation,  the
two mode period doubled limit cycle is only weakly stable and noise 
will be able to excite low frequency antiphase
dynamics. This explains the relatively broad region of antiphase
dynamics observed experimentally in Fig.~\ref{pra2}.   As we further
lower the detuning frequency to $\Delta \omega\approx 7\text{ GHz}$,
the two mode limit cycle exchanges stability with a single mode limit
cycle in a transcritical bifurcation (striped blue line in
Fig.~\ref{pra8}). At this point the uninjected mode is completely
switched off and the limit cycle in the single mode manifold becomes
stable, forming the lower boundary of the region (i). This scenario
is in agreement with experimental data of Fig. \ref{pra2}, where 
we observe that the spectral density evolves continuously across
the lower boundary of region (i). 

As we decrease $\Delta \omega$ further, the single mode period doubled
limit cycle undergoes an inverse period doubling bifurcation (dashed 
magenta line in Fig.~\ref{pra8}) and a single period limit cycle is 
generated. This is experimentally verified by the disappearance of the 
frequency at $\Delta \omega \approx 4 \text{ GHz}$, which happens
below the lower boundary of region (i), i.e. \textit{after} we have
switched to a single mode state. 

The generated single period limit cycle, which experimentally corresponds 
to the single mode wave mixing state, remains stable until we encounter 
a saddle node bifurcation of limit cycles (orange line SNL in Fig.
~\ref{pra8}) at $\Delta \omega\approx 4\text{ GHz}$. At this point the 
stable single mode limit cycle collides with an unstable one, and both 
limit cycles disappear \cite{kuz_95}. The dynamics are then blown out from 
the single mode manifold where the associated bursting dynamics explains 
the brief broadening which is observed in the experimental spectrum 
(Fig. \ref{pra2}) and in the numerical spectrum (Fig. \ref{pra5a}).
Close to the bifurcation point, the remainder of the stable limit cycle 
still forms a slow region on the single mode manifold, and thus the 
period of the resulting two mode orbit diverges close to the bifurcation 
point. Away from the region of bursting dynamics this new 
incommensurate frequency results in a quasiperiodic orbit on a torus
(cf. Fig \ref{pra7}). 

Moving away from the SNL bifurcation point to lower $\Delta \omega$, the 
frequency originating at zero increases linearly, and gives rise to satellite
peaks around the second major frequency close to the relaxation
oscillation frequency. This explains the striking star-like features
in the  experimental (Fig.~\ref{pra2}) and numerical (Fig.~\ref{pra5a}) 
spectra, which immediately follow the brief broadening discussed in the 
previous paragraph. Just below the SNL bifurcation, we are in a region
with two incommensurate frequencies, which gives rise to quasiperiodic
evolution on a stable torus manifold. The torus transforms into a limit cycle 
via an inverse torus bifurcation at $\Delta \omega \approx 3\text{ GHz}$ (red
line in Fig.~\ref{pra8}), and in the power spectrum, the low frequency that
was introduced by the SNL bifurcation disappears again.

Further lowering $\Delta \omega$ below $3\text{ GHz}$ we find that the
stable two mode limit cycle undergoes a period doubling cascade to
chaos. In Fig.~\ref{pra8} we only show the first period doubling
bifurcation (magenta line). At the transition to chaotic behaviour the
power spectrum broadens dramatically, which marks the transition from
region (ii) to region (iii). We leave the chaotic regime at $\Delta
\omega \approx 0$ in a single mode periodic orbit, which undergoes
inverse period doubling bifurcations (dashed magenta line in
Fig.~\ref{pra8}). 

As in basic single mode laser theory, this limit cycle becomes a
single mode equilibrium locked state (SL) via a Hopf bifurcation 
(dashed green line in Fig.~\ref{pra8}), which disappears in a saddle 
node bifurcation (dashed black line in Fig.~\ref{pra8}). 
Then the remaining single mode limit cycle undergoes again a SNL bifurcation 
followed by a torus bifurcation, as was discussed in the context of region (ii). 
This defines the two mode region (iv) of Fig.~\ref{pra1}. 

For even lower $\Delta \omega$  we find a two mode equilibrium state (TME), 
which is bounded by a Hopf bifurcation. This region corresponds to
the experimentally observed region below region (iv) in Fig.~\ref{pra1}, where the 
injected mode is suppressed. This suppression is a precursor to the appearance of 
a bistability between single-mode injection locking and a two mode equilibrium at 
larger negative detunings ($< -10$ GHz) and larger values of the injected field strength. 
We have recently demonstrated that this bistability can be the basis of an all-optical 
memory  element based on switching of the uninjected mode \cite{osborne_08}.  
Interestingly, there is also a second  two mode equilibrium region for positive 
detuning and weak injection ($0 < \Delta\omega < 6 \text{ GHz}, K < 0.005$) as shown 
in Fig. \ref{pra8}. This region will be the subject of more detailed study in future 
work.

\section{CONCLUSIONS}

We have presented an experimental and theoretical study of antiphase dynamics 
in an optically injected two-mode laser diode. The device was a specially 
engineered Fabry-Perot laser designed to support two primary modes with a 
large (THz) frequency spacing. At a fixed injected field strength, injection 
in one of the primary modes of the device leads to a rich dynamical scenario 
where dramatic switching between regions of two-mode and single mode dynamics 
was observed as the frequency detuning was varied. Using a minimal 
extension of the single mode rate equation model that includes a single 
equation for the carrier density and a general cross and self-saturation of 
the gain, we reproduced experimental results with a high degree of accuracy. 
In particular, we were able to describe the dynamical sequence of the total 
intensity and of the individual modes. These dynamics were shown 
to be organised by complex bifurcation structure that results from the 
increased dimensionality of the two-mode injected system. 

\textit{Acknowledgments.} This work was supported by Science Foundation Ireland 
and IRCSET. The authors thank Eblana Photonics for the preparation of sample
devices and G. Bordyugov for helpful discussions.

\end{document}